\documentclass[review]{elsarticle}
\usepackage[letterpaper, total={6in, 8in}]{geometry}

\usepackage{amssymb}
\usepackage{amsthm}

\usepackage{ulem}

\usepackage[figuresright]{rotating}

\usepackage{hyperref}
\hypersetup{
    colorlinks=true,
    linkcolor=blue,
    filecolor=magenta,      
    urlcolor=cyan,
}
\usepackage{float}
\usepackage{caption}
\usepackage{subcaption}

\begin{document}
\begin{frontmatter}
\title{Rapid screening of high-throughput ground state predictions}
\author{Sayan Samanta\corref{cor1}}
\ead{sayan\_samanta@brown.edu}
\author{Axel van de Walle}
\cortext[cor1]{Corresponding author}
\address{Box D, School of Engineering, Brown University, Providence, RI  02912-9104, USA}
\begin{abstract}
High-through computational thermodynamic approaches are becoming an increasingly popular tool to uncover novel compounds. However, traditional methods tend to be limited to stability predictions of stoichiometric phases at absolute zero. Such methods thus carry the risk of identifying an excess of possible phases that do not survive to temperatures of practical relevance. We demonstrate how the Calphad formalism, informed by simple first-principles input can be simply used to overcome this problem at a low computational cost and deliver quantitatively useful phase diagram predictions at all temperatures. We illustrate the method by re-assessing prior compound formation predictions and reconcile these findings with long-standing experimental evidence to the contrary.
\end{abstract}

\end{frontmatter}


\section{Introduction}

Phase diagram determination is a well-documented bottleneck for novel materials discovery and design \cite{phasediagDef}. To address this, high-throughput calculation tools \cite{Saal2013,Curtarolo2013,Morgan_2004} have been designed to provide
robust and automated end-to-end first principles pipelines to generate phase stability information.
However, traditional approaches in this field tend to focus solely on ordered stoichiometric phases, which entails important limitations \cite{avdw:mrspd}.
First, real engineering alloys tend to specifically exploit deviations from stoichiometry to optimize materials properties.
Second, a focus on ordered phases essentially implies that 
only phase stability at absolute zero is considered\footnote{although phonon contributions are at least  being increasingly considered in high-throughput settings \cite{toher:qdebye}}. This focus has the undesirable side-effect that it generally predicts a wide range of possible ordered ground states that exaggerate the number of phases that would typically observed at commonly accessible temperatures.


In this paper, we seek to address this last shortcoming by providing a simple way to verify whether an ordered phase predicted to be stable at absolute zero, based on high-throughput calculations, is indeed a phase that should be considered in a phase diagram covering room temperature and above.
Part of the solution is to leverage the CALculation of PHAse Diagrams (CALPHAD) \cite{Liu2009,Spencer20081,Sundman1981297} framework and the Compound Energy Formalism (CEF) \cite{hillert2001161,kusoffsky2001549,firsk2001177}.
Ab-initio electronic structure methods \cite{dft1,bigdeli201979} are routinely included to augment experimental input in CALPHAD assessments, \cite{Chen201566,Ong2019143,Arroyave2006473} but
these efforts typically focus on ordered phases. The inclusion of ab initio data for nonstoichiometric phases is less frequently attempted  \cite{Adjaoud,ghosh:sqsce,ghosh:hfnb,ghosh:AlZnTi} but when it is, it typically relies on the cluster expansion formalism, which is not easily amenable to a high-throughput treatment.

Methods that are more amenable to a high-throughput treatment in the construction of accurate phase diagrams at nonzero temperature constructed purely computationally from first principles could revolutionise the process of new materials screening and discovery in an unprecedented manner \cite{mediukh2019101643,wang201955,burton2006222}. In this work we demonstrate the effectiveness of a software pipeline that can model the thermodynamics of an alloy system by fitting the coefficients of a polynomial expression representing excess solution free energies in the CALPHAD formalism to a set of distinct ab-initio calculation  \cite{sqs2tdb}. To expedite the ab-initio calculations, this framework relies on the concept of Special Quasirandom Structure (SQS)\cite{zunger.65.353,mcsqs201313}. 
An SQS is a representation of a completely disordered alloy that attempts to mimic the short-range correlations functions of a {\em true} completely disordered material in a small periodic simulation cell. The code \cite{sqs2tdb} generates select compositions which are a function of the phase being modeled and the number of components from a SQS database.
A key distinguishing feature of this software system is that it offers an independent control over the different sublattice compositions (i.e., site fractions) when exploring structure space, which allows the determination all the CEF parameters from the ab initio data (and not only the end member energies).
Once the energetics of each configuration is completed, the code then packages the result in a Thermodynamic Database (TDB) file that can then be used by standard CALPHAD modeling tools \cite{thermoCalc1985153,FactSage2002189,FactSage2009295,pandat2009328,OpenCalphad2015} for further analysis.

For the aforementioned demonstrations, we chose to investigate the binary Iridium-Ruthenium phase diagram. Iridium and Ruthenium (alloys and oxides), have primarily been of interest in the catalysis community and has applications in treatment of breast cancer \cite{wei2020119848}, coatings  \cite{vukovic1992663}, oxygen evolution reaction \cite{shan2019445,Ktz_1985}. Hart et. al. \cite{hart:PGM} also published the results of a large number of stable binary phases in alloys of the Pt-group metals (PGMs), in which it was reported that Ir-Ru binary alloys may have possible stable ordered phases. Thus, it is a good candidate alloy system for our illustration.

\section{Methods}

The process of quickly generating a phase diagram from ab initio data has been explained in detail in reference \cite{sqs2tdb} and illustrated in \cite{avdw:resubpd}. The theoretical framework and the description of each command and of the code can be found in that paper as well. Here, we only outline the major steps of the process for the specific example system considered.
\begin{enumerate}
\item Setting up structures for electronic structure calculation: 

Sampling the accessible composition range of each potential phase using pre-generated Special Quasirandom Structures (SQS). Here, based on the high-throughput study from \cite{hart:PGM}, we consider the hcp, fcc, L1$_2$ \cite{l122004635} and D0$_{19}$ \cite{D019Kang2001} phases, to which we add the liquid phase. We jointly model the fcc and L1$_2$ as well as the hcp and D0$_{19}$ phases using the same CEF model.
The composition grid  used includes samples of each sublattice compositions in steps of 50 \% with additional point at 25 \% interval on either one sublattice when the other sublattice is defect-free. This corresponds to a composition sampling density ``level'' of 3 in the {\verb sqs2tdb } structure generation command (see Appendix for details). For the sake of completeness we also have included the other phase (B19, Pt$_8$Ti, Ir$_2$Tc, Hf$_5$Sc) modeled as line compounds even though we had strong indications that those phases will not be energetically competitive. We confirmed this exclusionary hypothesis through our calculations as well.
\item Calculating ab initio energies for each structure:

One can calculate the electronic ground state formation energies of each of the structures using any electronic structure code. In this work, we used the plane wave basis projector augmented wave method (PAW) \cite{PAW_PhysRevB.50.17953} with the exchange correlation approximation being the generalised gradient approximation (GGA) in the Perdew-Burke-Ernzerhof (PBE) \cite{PBE_PhysRevLett.77.3865} form as implemented in the Vienna Ab-initio Simulation Package (VASP) \cite{vasp_kresse199615,vasp_PhysRevB.54.11169}. All calculations were conducted at a precision flag of \textit{Accurate} (to set the plane wave cut-off as the maximum of the ENMAX in the POTCAR of the participating species) and an ionic convergence criterion of $10^{-4}$ eV. The optimisation was performed using the conjugate gradient algorithm \cite{CG10.5555/148286}. The no. of kpoints to sample the first Brillouin Zone was generated automatically at a fixed density of $20^3 \AA^{-3}$. The partial electronic occupancies were set according to the Methfessel-Paxton scheme of order 1 \cite{MPPhysRevB.40.3616} with a smearing width of 0.1 eV (for overall good force and energies). At the end of the initial relaxation run, a further static electronic relaxation was performed using the tetrahedron method with Bl\"{o}chl correction \cite{BlochlPhysRevB.49.16223} (for more accurate energy calculation of the ground state configuration)

Ab-initio molecular dynamics \cite{AIMDK_hne_2014} was run at temperature above the melting point of Ir and Ru to obtain the energy of the liquid phase. The MD runs were conducted at fixed number of atoms, fixed volume and temperature (NVT ensemble) employing a Langevin Thermostat \cite{Langevindoi:10.1063/1.445195,LangevinPhysRevLett.48.1818}.
The desired external pressure of 0 was achieved by adjusting the simulation cell size.
Calculation settings (potential and exchange correlation functional, cutoffs, k-points, etc.) were kept consistent with that of the relaxation runs, except for the Fermi smearing, which was set to match the ionic temperature of 3000 K.  MD was run for 3300 steps, with each step size being 1.5 fs for a total time of 4.95 ps. We considered the first 1000 steps as equilibration and averaged the energy of the final 2300 steps. The liquid composition sampling grid was kept coarse (levels 0 and 1 in {\verb sqs2tdb }).

\item  Calculating Vibrational Entropies:

To calculate the effect of atomic vibrations in the neighborhood of the relaxed configuration at finite temperature, the \verb fitfc  code was used \cite{fitfc_2009}. Symmetrically distinct configurations were generated by displacing atoms from their relaxed position and calculating the static energies of these configurations. Subsequently the reaction forces induced by the displacements (frozen in place) were fit to a harmonic spring model to calculate the free energy of vibrations. This standard procedure is elaborated in \cite{fitfc_2009}.
In practice, a good speed/accuracy trade-off is obtained by performing these calculations for endmembers only (corresponding to a composition sampling accuracy ``level'' of 0). However, more accurate phase boundaries might demand similar calculations on a finer compositions sampling.

In the present work, phonon vibrations were calculated for the end-members of all phases and additionally up to level 1 for the fcc and hcp phase.
 
In all calculations, the minimum distance between periodic images of the displaced atoms were kept to 4 times the nearest neighbor distance for the phase under considerations (corresponding to the setting {\verb frnn=4 } in {\verb fitfc } ), while the range of the springs included while fitting a purely harmonic model ({\verb ns=1 } setting in {\verb fitfc }) was at 2 times the nearest neighbor distance ({\verb ernn=2 } setting in \verb fitfc) . The magnitude of displacement of the atoms from the relaxed state were kept at {\verb dr=0.04 } \AA.

In cases where there were spurious mechanically unstable phonon modes, the procedure as described in the paper \cite{fitfc_2009} was adopted, in which the code generates displacements along the unstable directions to double-check stability. If the instability is not confirmed, the newly generated reaction forces are added to the fit in order to correct the problem.

\item Reference states used:

All formation energies were referenced to fcc Ir and hcp Ru. We did not make use of the experimental SGTE free energy difference between the fcc and hcp pure phases. The reason for proceeding in this way is that the SGTE and DFT hcp-fcc free energy differences do not agree very well \cite{vandewall_SGTE20181} and we wanted our results to purely reflect the DFT predictions. The corresponding difference for Ru does agree well, so our strategy has little effect on the Ru-rich side of the phase diagram. In the {\verb sqs2tdb } code, this choice of reference states is indicated by including the hcp Ir and fcc Ru phases in the file {\verb exfromsgte.in } in the parent folder of the phase diagram calculations. It should be noted that hcp Ir and fcc Ru are not mechanically unstable, so the discrepancies with SGTE values found here can be attributed to DFT and not any specific treatment of mechanical instability \cite{mech_vandeWalle2015}.

\item Fitting the data with CALPHAD model:

With all the energetics computed, a least square fit is performed to obtain the coefficients of the different polynomial thermodynamic functions (formation energy and vibrational formation free energy) according to the CALPHAD formalism.
The numbers of terms in the excess free energy function for binary interactions in a sub-lattice of the phase under consideration was limited to 2. The inherent constraint of the CALPHAD formalism dictates that only one sublattice can have higher order interactions per term.
\end{enumerate}

It should be noted that in this system, the predicted L1$_2$ and D0$_{19}$ phases from \cite{hart:PGM} turn out to be superstructures of fcc and hcp, respectively. This suggests making use of a multiple-sublattice framework to jointly handle each ordered phase and its associated disordered phase in a single phase description \cite{kusoffsky2001549,ansara199720}. This approach enables a good approximation of short-range-order (SRO) effects in the solid solution phases by allowing the different sublattices to adopt slightly different compositions. While it would be admittedly preferable to have fcc and hcp phases with uniform sublattice-independent compositions in a fully optimized CALPHAD model of the system, the simple scheme adopted here is ideal for rapid screening. As an input, this approach only demands a few extra SQS calculations in which the different sublattices have different compositions rather than necessitating the use of more advanced models, such as a cluster expansion.

\section{Results and Discussion}

\begin{figure}[H]
\centering
\includegraphics[width=\textwidth]{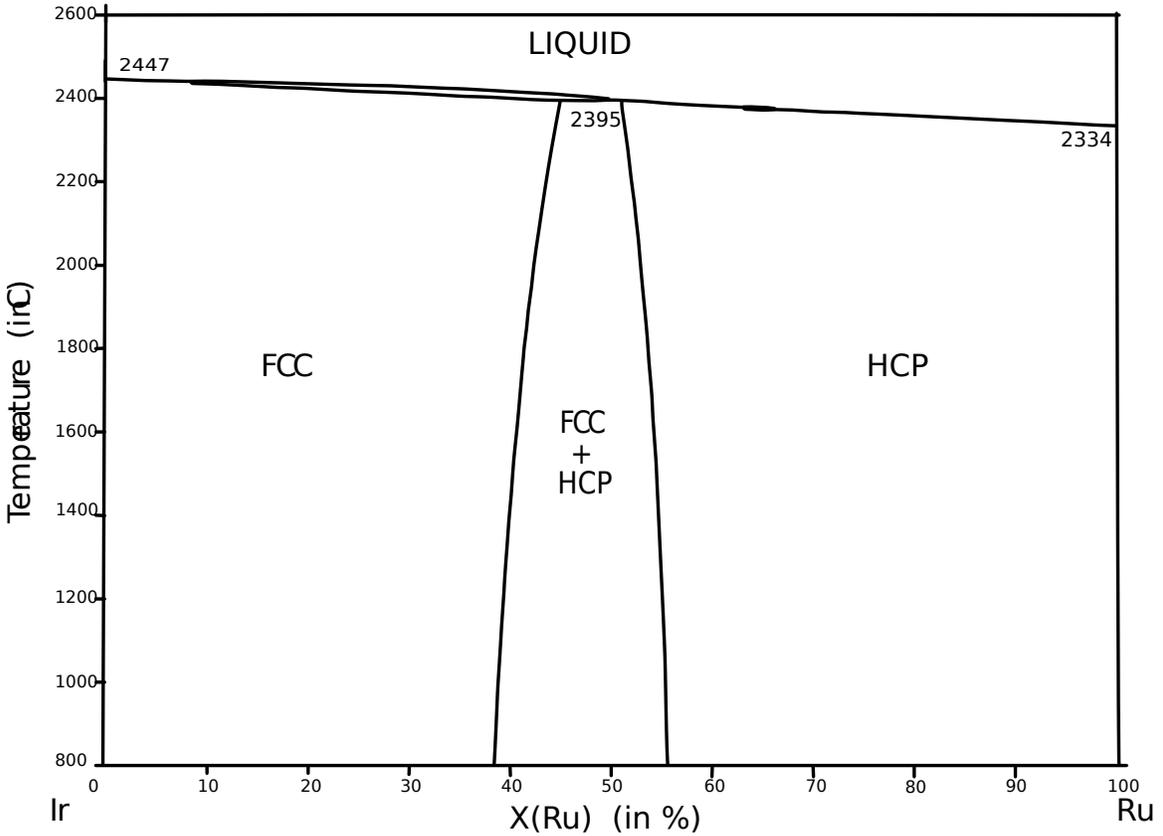}
\caption{Experimental phase Diagram for Ir-Ru binary alloy, Okamoto H., Ir-Ru (Iridium-Ruthenium), Binary Alloy Phase Diagrams, II Ed., Ed. T.B. Massalski, Vol. 3, 1990, p 2345-2348, adapted from ASM International Alloy Phase Diagram database \cite{okamoto_pd}}
\label{og_pd}
\end{figure}

\begin{figure}[H]
\centering
\includegraphics[width=\textwidth]{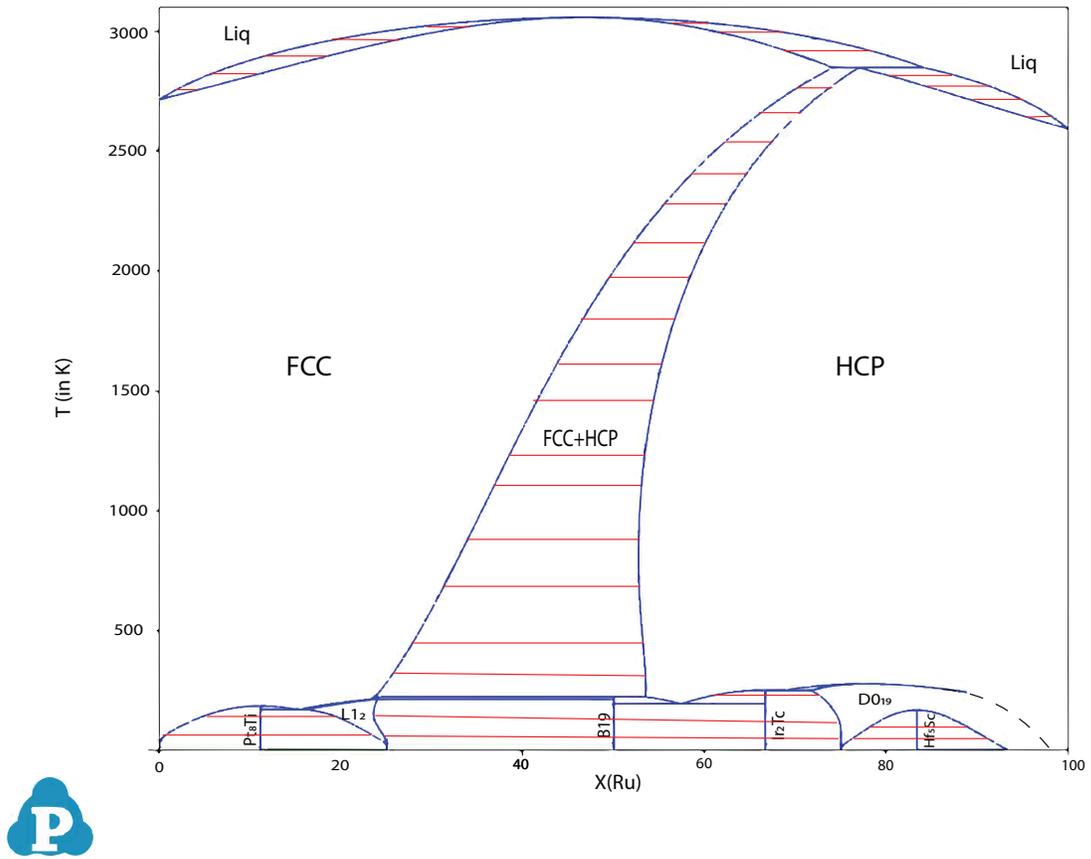}\label{ordered_phases}
\caption{Calculated Phase Diagram}
\label{pd}
\end{figure}

The prediction by Hart and colleagues \cite{hart:PGM} of new ordered phases in the Ir-Ru system is surprising, given that existing experimental assessments of this system find no such phases \cite{IrRuphase1,IrRuphase2,Ktz_1985}
(see, for instance, Figure \ref{og_pd}). Our calculated Ir-Ru phase diagram (reported in Figure \ref{pd}) immediately resolve this apparent conundrum: all predicted ordered phases disorder below room temperature, making it unlikely that such phases could form through standard heat treatments.However, as our method does not account for SRO within each of the sublattices of L1$_2$ and D0$_{19}$, there is a risk that our calculated free energies for those phases are biased upward.

\begin{figure}[H]
\centering
\includegraphics[width=\textwidth]{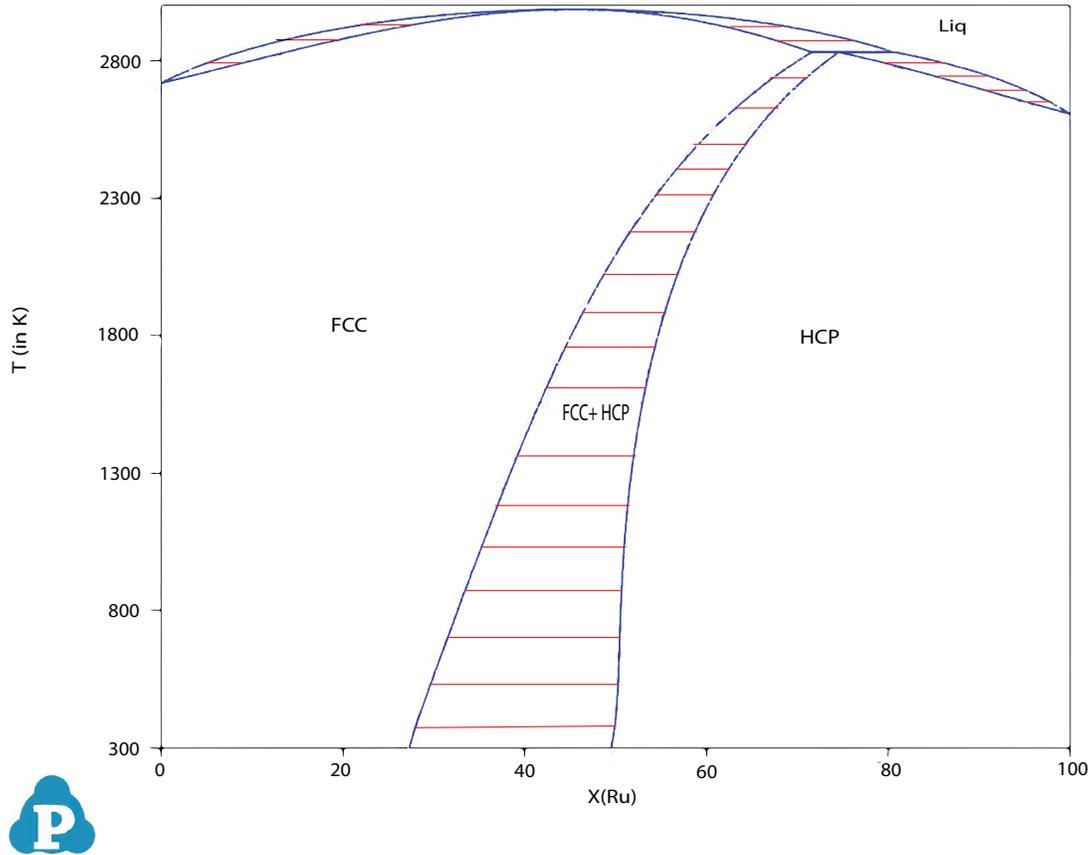}\label{fcc-hcp-liq}
\caption{Calculated Phase Diagram (with FCC, HCP and Liquid only)}
\label{pd_fcc-hcp-liq}
\end{figure}
Our calculated liquidus and solidus do not agree particularly well with the experimental assessments, but this is expected given the relatively simplified treatment used for the liquid state (only formation energies are calculated while entropic contributions solely come from the end members' entropy). Likewise, the shift in the predicted
fcc-hcp two-phase region at high temperature, relative to the experimental result, is likely due to the use of a harmonic approximation the phonon calculations that becomes progressively less justified at higher temperatures. These inaccuracies, however, are inconsequential for our finding, namely that the predicted ordered phases become unstable at a relatively low temperature.

\begin{figure}[H]
\centering
\begin{subfigure}[b]{0.8\textwidth}
\includegraphics[width=\textwidth]{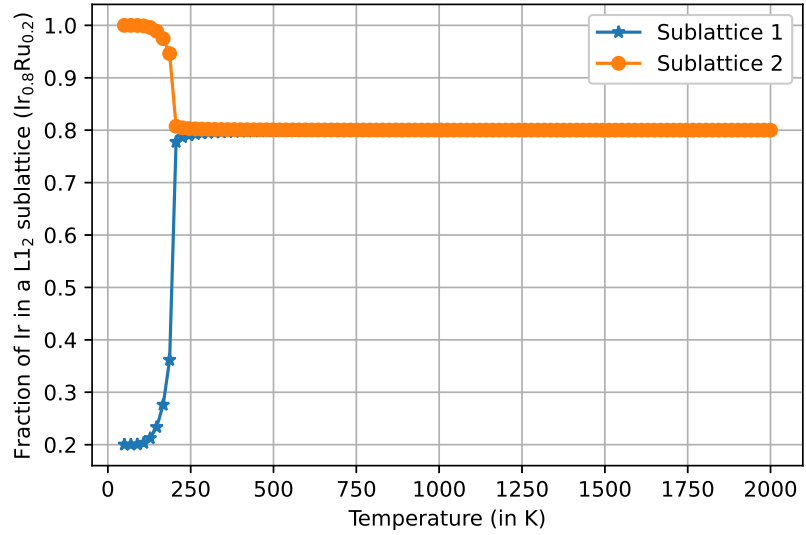}
\caption{}
\end{subfigure}
\begin{subfigure}[b]{0.8\textwidth}
\includegraphics[width=\textwidth]{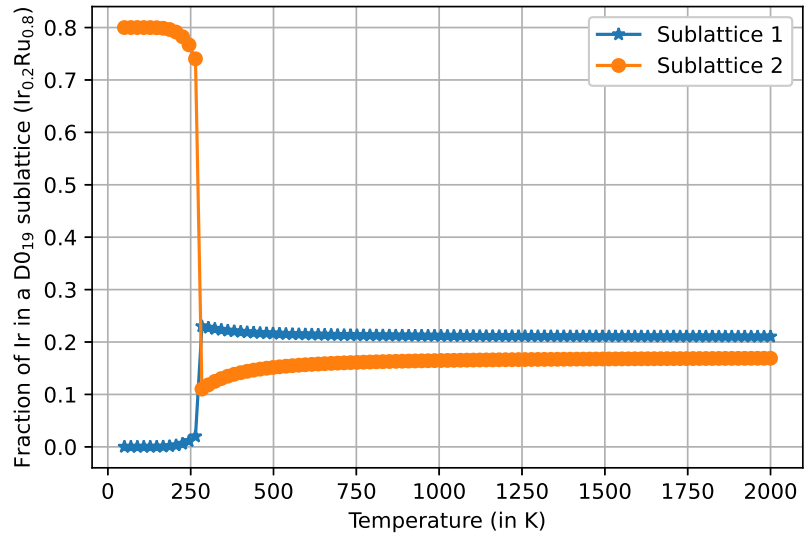}
\caption{}
\end{subfigure}
\caption{(a) Distribution of Iridium in each sublattice of L$1_2$ at an overall composition of Ir$_{0.8}$Ru$_{0.2}$, , (b) Distribution of Iridium in each sublattice of D0$_{19}$ at an overall composition of Ir$_{0.2}$Ru$_{0.8}$, across a range of temperature 50-2000 K}
\label{break_ordered}
\end{figure}

Even though disordering occurs at low-temperatures in this system, the predicted ground states could still provide some information regarding the nature of the SRO in this system. To gain further insight into the system's ordering behavior, we calculate the equilibrium sublattices compositions (i.e. site fractions) for each of the ordered phases as a function of temperature (see Figure \ref{break_ordered}). We select overall compositions such that the graphs cut through the point of highest temperature in the phase's respective region of stability. For both the L1$_2$ and D0$_{19}$ phases, we find that the sublattice compositions sharply change at the transition temperature, consistent with a first order transition. The fact that the two sublattices of L1$_2$ have the essentially the same composition above the transition suggests that the fcc phase exhibits little SRO. For the D0$_{19}$ phase, the small ($<$ 5\%) remaining difference in sublattice compositions above the transition temperature suggests that SRO in the hcp phase is also small, although more pronounced than in the fcc phase.

Given that the decomposition temperatures of all the ordered phases are so low, it proves more convenient for most applications to have a thermodynamic model that ignores the low temperature ordered phases and provides a single-sublattice description of the fcc and hcp phases. We have generated a such a thermodynamic model, using standard single-sublattice SQS and including the effect of short-range order using the Cluster Variation Method-based scheme described in \cite{sqs2tdb}. This database is included in the supplementary information and the corresponding phase diagram is shown in Figure \ref{pd_fcc-hcp-liq}.

\begin{figure}[H]
\centering
\begin{subfigure}{0.8\textwidth}
\includegraphics[width=0.8\textwidth]{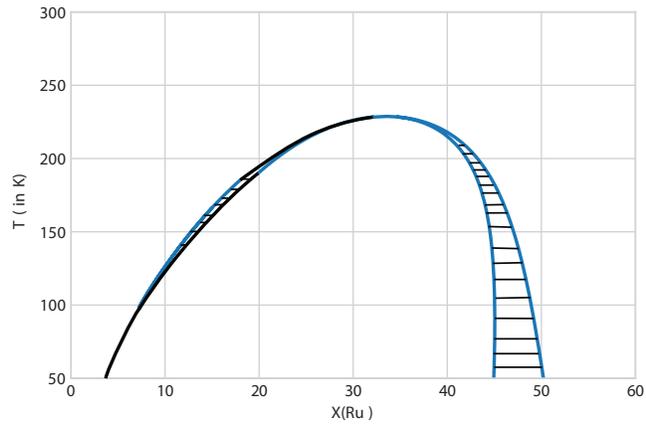}
\caption{}
\label{fig:l12_metastable}
\end{subfigure}
\begin{subfigure}{0.8\textwidth}
\includegraphics[width=0.8\textwidth]{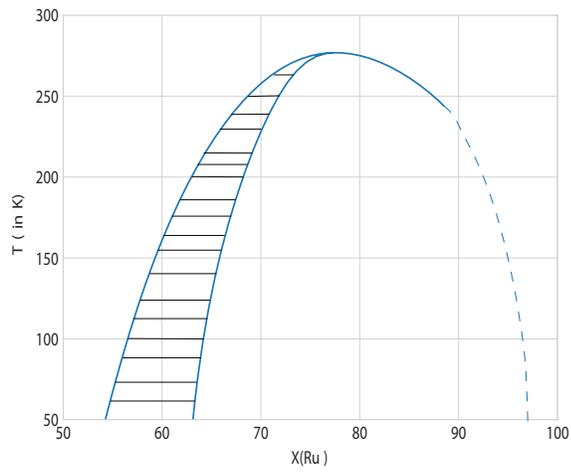}
\caption{}
\label{fig:d019_metastable}
\end{subfigure}
\caption{(a) L1$_{2}$, (b) D0$_{19}$ phases only between 50-300 K, dashed lines denote a possible second order phase transition}
\label{fig:singlephase_lowT}
\end{figure}

\begin{figure}[H]
\centering
\begin{subfigure}{0.8\textwidth}
\includegraphics[width=\textwidth]{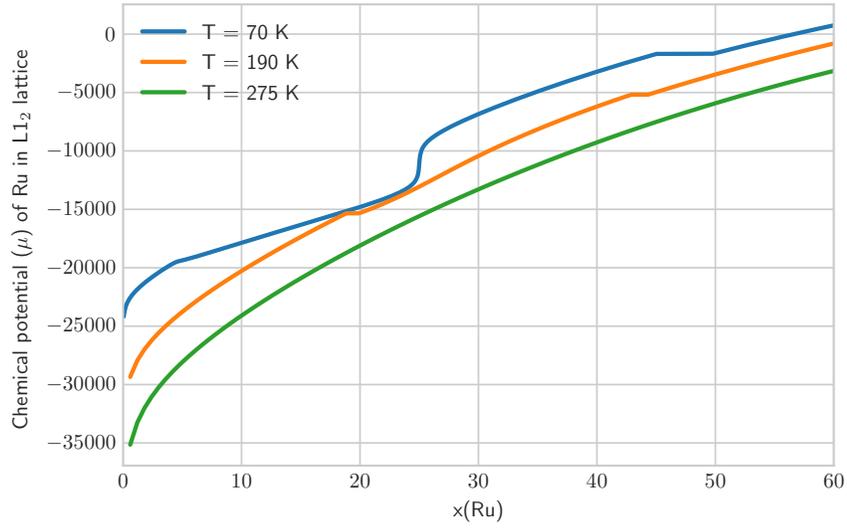}
\caption{}
\label{fig:l12_chempot}
\end{subfigure}
\begin{subfigure}{0.8\textwidth}
\centering
\includegraphics[width=\textwidth]{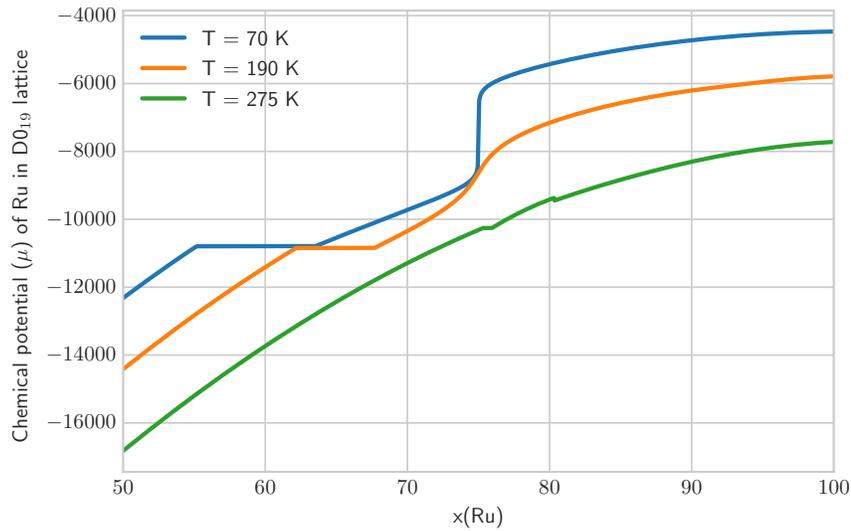}
\caption{}
\label{fig:d019_chempot}
\end{subfigure}
\caption{(a) Chemical potential ($\mu$) of Ru in L1$_2$ over a composition range of 0-60\% Ru for T = 70, 190, 275 K, (b) Chemical potential ($\mu$) of Ru in D0$_{19}$ over a composition range of 50-100\% Ru for T = 70, 190, 275 K}
\label{chempots}
\end{figure}

It is also instructive to study in more detail the order-disorder transition by considering the metastable phase diagram for fcc-based and hcp-based phases separately (see Figure \ref{fig:singlephase_lowT}).
At the level of accuracy of our thermodynamic description we use, it is not possible to unambiguously identify second-order transitions, as all parametrizations used are smooth in composition and temperature by construction. However, whenever the width of a two-phase region shrinks to a point, this is suggestive that the phase transition becomes second-order. The L1$_2$ phase appears to be surrounded by first-order transitions, albeit very narrow ones on the Ir-rich side. The Ru-rich side transition are masked by hcp phases in the full phase diagram. The D0$_{19}$ phase exhibits clear first-order transition towards the Ir-rich side while the first-order character completely vanishes towards the Ru-rich side. To ascertain that this disappearance of the first-order transition is not merely a numerical artifact, we also plotted the chemical potential as a function of composition in Figure \ref{chempots}. These plots clearly confirm the absence of first-order transition on the Ru-rich side of the D0$_{19}$ phase, which would have been seen as a horizontal line. We conjecture that the phase transition becomes second-order and we indicate this possibility by a dashed line in Figure \ref{fig:d019_metastable} (although its exact position cannot be determined from the present calculations).

\section{Conclusion}

Our first principles approach to the construction of phase diagram, applied to the Iridium-Ruthenium Binary alloy system, has successfully helped explain certain inconsistencies between past experimental and high-throughput results. The absence of stable ordered phases, (such as the L1$_2$ and D0$_{19}$ phases in this work), in contrast to the predictions from high-throughput calculations, can be attributed to order-disorder transitions at a temperature range lower than the temperature range at which experimental phase diagrams are constructed. 

In the process, we observe that representing disordered phases by a partially ordered multi-sublattice can act as a good proxy short range ordering (SRO) behavior. \textcolor{blue}{This is advantageous as SRO is computationally expensive \cite{sro:kikuchi,sro:Pelizzola_2005} to explicitly evaluate, although alternative KKR-CPA-based approaches \cite{sro:kkr-cpa1,sro:johnson,sro:cpa_soven,sro:Zeller_1987}, have been shown to be helpful to address this}. In general, the CALPHAD formalism, in conjunction with our technique, can provide a general representation for both solid solutions and stoichiometric phases that is amenable to a high-throughput pipeline. 

We also observe that entropic contributions to the free energy at higher temperatures can be approximated to a reasonable degree of accuracy by phonon calculations on a few structures (often end-members). For further improvements (particularly to address the greater mismatch solidus and liquidus curves between our calculation and experiments at higher temperature), we postulate that adopting anharmonic spring models and/or denser compositional grids would result in better estimations. A better model of the entropy of the liquid would also likely be needed.

For the purpose of speeding up the liquid calculations, our group is also working on including the liquid phase within the SQS framework. Our method nevertheless provides a automated pipeline for quick assessment of phase diagrams for exploratory purposes, that can be incrementally improved at each step of the calculation to desired levels of accuracy. Although the process described is with first principles calculations only, it can be easily augmented, when fitting the CALPHAD model, with experimental data to further enhance the accuracy.

\section*{Acknowledgements}

The authors acknowledge support from the National Science Foundation through grant DMR-2001411. Computational resources were provided by the Center for Computation and Visualization at Brown University and the Extreme Science and Engineering Discovery Environment (XSEDE) Stampede2 at the Texas Advanced Computing Center through allocation TG-DMR050013N, which is supported by National Science Foundation Grant No. ACI-1548562.

\section{Data Availability}

The raw data required to reproduce these findings can be obtained by following the steps as illustrated in \ref{appendix:atat}
The processed data required to reproduce the phase diagrams and the supporting findings are available to download from this \href{https://github.com/reach2sayan/Iridium-Ruthenium-Binary-Thermodynamic-Database}{link} and have also been provided as supporting files with this article.

\appendix
\section{ATAT Commands used in this work}
\label{appendix:atat}
To facilitate reproducing the results and help readers undertake their own analysis on other systems, we here provide the commands of the ATAT package used in this work (we omit ab initio code-specific commands).

\begin{enumerate}
    
\item To generate the input structures for the ab initio calulations, we use:
  
\begin{verbatim} sqs2tdb -cp -lv=[1,2,3], -l=[FCC_A1,BCC_A2,...] -sp=Ir,Ru \end{verbatim}
  
This creates a set of folder for each phase where the SQS structures up to the desired levels are copied. We can run our ab-initio code inside these to calculate the electronic ground state structure and energy.

\item For the phonon calculations:

For every level of a particular phase for which we want to include phonon vibrational entropy calculations, we can generate symmetrically distinct configurations with specified displacements per atom and distance between the periodic images of the displaced atom (here the distance between two periodic atom images is in units of nearest neighbour distance. Check \cite{fitfc_2009} for more details).

\begin{verbatim} fitfc -ernn=4 -ns=1 -dr=0.04 \end{verbatim}

This creates a set of folders \begin{verb} vol_*/p* \end{verb} representing every symmetrically distinct perturbed structures.  Calculate energy of each from ab-initio method. Then we can fit the forces to a specified spring model (here {\verb ns=1 } indicate a harmonic model), using the code snippet

\begin{verbatim} fitfc -frnn=2 -ns=1 -dr=0.04 -fu \end{verbatim}

The {\verb -fu } tag provides information on the unstable modes if any suggested by the initial fit report suggestive of the presence of mechanical instability. If any such modes are suspected, we can read the file \begin{verb} vol_*/unstable.out \end{verb}. Re-running the fitting procedure with an addition tag \begin{verb} -gu=[index] \end{verb}, creates a supercell representing the mode describing the imaginary mode which is described in the row matching the \begin{verb} [index] \end{verb} at the second column of the \begin{verb} vol_*/unstable.out \end{verb} file . We can then calculate the reaction forces in the direction and include it in the fit. Often if the instability is an artifact of the fitting process (imaginary modes), this process can solve the issue. Note that this process might have to be repeated multiple times until we include all the suspected unstable modes in our fit or unless a true unstable mode is found.

\item Generate Thermodynamic Database file:

We first input to the code the types of interactions (just single atom, binary, ternary etc.) and to which levels (0,1,2 etc.) in the excess free energy term as per the CALPHAD formalism \cite{Sundman1981297} in to the {\verb terms.in } file with the following format

\begin{verbatim}cat <<EOF > terms.in
1,0:1,0
2,1:1,0
EOF
\end{verbatim}

(the example shown is for a two sublattice phase where we include all first order terms (only level 0 is possible). We also include order 2 terms with 1 level. Note. that the CALPHAD model only allows greater than 1 order for only one sublattice at a time.)

Then run the command:
    
\begin{verbatim} sqs2tdb -fit [-sro] \end{verbatim}
 
This creates a partial Thermodynamic DataBase (TDB) file containing the function and structure for the particular phase. The optional parameter {\verb -sro } can be specified to include short-range order effects in single-sublattice phases (this was used for the fcc and hcp phases in this work). It is at this step when the code decides which reference values are to be drawn from SGTE database or not. These individual TDB files (one for each phase) can be combined in to a single TDB file for the alloy system using the command.

\begin{verbatim} sqs2tdb -tdb [-oc] \end{verbatim}

The {\verb -oc } tag is optional used to create the TDB file in the OpenCalphad format (OC) \cite{OpenCalphad2015}

\end{enumerate}



\bibliographystyle{elsarticle-num}
\bibliography{bibliography.bib}

\begin{thebibliography}{10}
\expandafter\ifx\csname url\endcsname\relax
  \def\url#1{\texttt{#1}}\fi
\expandafter\ifx\csname urlprefix\endcsname\relax\def\urlprefix{URL }\fi
\expandafter\ifx\csname href\endcsname\relax
  \def\href#1#2{#2} \def\path#1{#1}\fi

\bibitem{phasediagDef}
\href{http://www.sciencedirect.com/science/article/pii/B9780444537706180013}{Index},
  in: D.~E. Laughlin, K.~Hono (Eds.), Physical Metallurgy (Fifth Edition),
  fifth edition Edition, Elsevier, Oxford, 2014, pp. 2837 -- 2899.
\newblock \href
  {http://dx.doi.org/https://doi.org/10.1016/B978-0-444-53770-6.18001-3}
  {\path{doi:https://doi.org/10.1016/B978-0-444-53770-6.18001-3}}.
\newline\urlprefix\url{http://www.sciencedirect.com/science/article/pii/B9780444537706180013}

\bibitem{Saal2013}
J.~E. Saal, S.~Kirklin, M.~Aykol, B.~Meredig, C.~Wolverton,
  \href{https://doi.org/10.1007/s11837-013-0755-4}{Materials design and
  discovery with high-throughput density functional theory: The open quantum
  materials database (oqmd)}, JOM 65~(11) (2013) 1501--1509.
\newblock \href {http://dx.doi.org/10.1007/s11837-013-0755-4}
  {\path{doi:10.1007/s11837-013-0755-4}}.
\newline\urlprefix\url{https://doi.org/10.1007/s11837-013-0755-4}

\bibitem{Curtarolo2013}
S.~Curtarolo, G.~L.~W. Hart, M.~B. Nardelli, N.~Mingo, S.~Sanvito, O.~Levy,
  \href{https://doi.org/10.1038/nmat3568}{The high-throughput highway to
  computational materials design}, Nature Materials 12~(3) (2013) 191--201.
\newblock \href {http://dx.doi.org/10.1038/nmat3568}
  {\path{doi:10.1038/nmat3568}}.
\newline\urlprefix\url{https://doi.org/10.1038/nmat3568}

\bibitem{Morgan_2004}
D.~Morgan, G.~Ceder, S.~Curtarolo,
  \href{https://doi.org/10.1088%2F0957-0233%2F16%2F1%2F039}{High-throughput and
  data mining with ab initio methods}, Measurement Science and Technology
  16~(1) (2004) 296--301.
\newblock \href {http://dx.doi.org/10.1088/0957-0233/16/1/039}
  {\path{doi:10.1088/0957-0233/16/1/039}}.
\newline\urlprefix\url{https://doi.org/10.1088%2F0957-0233%2F16%2F1%2F039}

\bibitem{avdw:mrspd}
A.~van~de Walle, M.~Asta, {{}}{H}igh-throughput calculations in the context of
  alloy design, MRS Bull. 44 (2019) 252.
\newblock \href {http://dx.doi.org/10.1557/mrs.2019.71}
  {\path{doi:10.1557/mrs.2019.71}}.

\bibitem{toher:qdebye}
C.~Toher, J.~J. Plata, O.~Levy, M.~de~Jong, M.~Asta, M.~B. Nardelli,
  S.~Curtarolo, High-throughput computational screening of thermal
  conductivity, {D}ebye temperature, and {G}r{\"{u}}neisen parameter using a
  quasiharmonic {D}ebye model, Phys. Rev. B 90 (2014) 174107.

\bibitem{Liu2009}
Z.-K. Liu, \href{https://doi.org/10.1007/s11669-009-9570-6}{First-principles
  calculations and calphad modeling of thermodynamics}, Journal of Phase
  Equilibria and Diffusion 30~(5) (2009) 517.
\newblock \href {http://dx.doi.org/10.1007/s11669-009-9570-6}
  {\path{doi:10.1007/s11669-009-9570-6}}.
\newline\urlprefix\url{https://doi.org/10.1007/s11669-009-9570-6}

\bibitem{Spencer20081}
P.~Spencer,
  \href{http://www.sciencedirect.com/science/article/pii/S0364591607000764}{A
  brief history of calphad}, Calphad 32~(1) (2008) 1 -- 8.
\newblock \href
  {http://dx.doi.org/https://doi.org/10.1016/j.calphad.2007.10.001}
  {\path{doi:https://doi.org/10.1016/j.calphad.2007.10.001}}.
\newline\urlprefix\url{http://www.sciencedirect.com/science/article/pii/S0364591607000764}

\bibitem{Sundman1981297}
B.~Sundman, J.~Ågren,
  \href{http://www.sciencedirect.com/science/article/pii/002236978190144X}{A
  regular solution model for phases with several components and sublattices,
  suitable for computer applications}, Journal of Physics and Chemistry of
  Solids 42~(4) (1981) 297 -- 301.
\newblock \href
  {http://dx.doi.org/https://doi.org/10.1016/0022-3697(81)90144-X}
  {\path{doi:https://doi.org/10.1016/0022-3697(81)90144-X}}.
\newline\urlprefix\url{http://www.sciencedirect.com/science/article/pii/002236978190144X}

\bibitem{hillert2001161}
M.~Hillert,
  \href{http://www.sciencedirect.com/science/article/pii/S092583880001481X}{The
  compound energy formalism}, Journal of Alloys and Compounds 320~(2) (2001)
  161 -- 176, materials Constitution and Thermochemistry. Examples of Methods,
  Measurements and Applications. In Memoriam Alan Prince.
\newblock \href
  {http://dx.doi.org/https://doi.org/10.1016/S0925-8388(00)01481-X}
  {\path{doi:https://doi.org/10.1016/S0925-8388(00)01481-X}}.
\newline\urlprefix\url{http://www.sciencedirect.com/science/article/pii/S092583880001481X}

\bibitem{kusoffsky2001549}
A.~Kusoffsky, N.~Dupin, B.~Sundman,
  \href{http://www.sciencedirect.com/science/article/pii/S036459160200007X}{On
  the compound energy formalism applied to fcc ordering.}, Calphad 25~(4)
  (2001) 549 -- 565.
\newblock \href
  {http://dx.doi.org/https://doi.org/10.1016/S0364-5916(02)00007-X}
  {\path{doi:https://doi.org/10.1016/S0364-5916(02)00007-X}}.
\newline\urlprefix\url{http://www.sciencedirect.com/science/article/pii/S036459160200007X}

\bibitem{firsk2001177}
K.~Frisk, M.~Selleby,
  \href{http://www.sciencedirect.com/science/article/pii/S0925838800014821}{The
  compound energy formalism: applications}, Journal of Alloys and Compounds
  320~(2) (2001) 177 -- 188, materials Constitution and Thermochemistry.
  Examples of Methods, Measurements and Applications. In Memoriam Alan Prince.
\newblock \href
  {http://dx.doi.org/https://doi.org/10.1016/S0925-8388(00)01482-1}
  {\path{doi:https://doi.org/10.1016/S0925-8388(00)01482-1}}.
\newline\urlprefix\url{http://www.sciencedirect.com/science/article/pii/S0925838800014821}

\bibitem{dft1}
R.~O. Jones, \href{https://link.aps.org/doi/10.1103/RevModPhys.87.897}{Density
  functional theory: Its origins, rise to prominence, and future}, Rev. Mod.
  Phys. 87 (2015) 897--923.
\newblock \href {http://dx.doi.org/10.1103/RevModPhys.87.897}
  {\path{doi:10.1103/RevModPhys.87.897}}.
\newline\urlprefix\url{https://link.aps.org/doi/10.1103/RevModPhys.87.897}

\bibitem{bigdeli201979}
S.~Bigdeli, L.-F. Zhu, A.~Glensk, B.~Grabowski, B.~Lindahl, T.~Hickel,
  M.~Selleby,
  \href{http://www.sciencedirect.com/science/article/pii/S0364591618301718}{An
  insight into using dft data for calphad modeling of solid phases in the third
  generation of calphad databases, a case study for al}, Calphad 65 (2019) 79
  -- 85.
\newblock \href
  {http://dx.doi.org/https://doi.org/10.1016/j.calphad.2019.02.008}
  {\path{doi:https://doi.org/10.1016/j.calphad.2019.02.008}}.
\newline\urlprefix\url{http://www.sciencedirect.com/science/article/pii/S0364591618301718}

\bibitem{Chen201566}
B.~Chen, Y.~Li, X.~Guan, C.~Wang, C.~Wang, Z.~Gao,
  \href{http://www.sciencedirect.com/science/article/pii/S0927025615002487}{First-principles
  study of structural, elastic and electronic properties of zrir alloy},
  Computational Materials Science 105 (2015) 66 -- 70.
\newblock \href
  {http://dx.doi.org/https://doi.org/10.1016/j.commatsci.2015.04.014}
  {\path{doi:https://doi.org/10.1016/j.commatsci.2015.04.014}}.
\newline\urlprefix\url{http://www.sciencedirect.com/science/article/pii/S0927025615002487}

\bibitem{Ong2019143}
S.~P. Ong,
  \href{http://www.sciencedirect.com/science/article/pii/S0927025619300138}{Accelerating
  materials science with high-throughput computations and machine learning},
  Computational Materials Science 161 (2019) 143 -- 150.
\newblock \href
  {http://dx.doi.org/https://doi.org/10.1016/j.commatsci.2019.01.013}
  {\path{doi:https://doi.org/10.1016/j.commatsci.2019.01.013}}.
\newline\urlprefix\url{http://www.sciencedirect.com/science/article/pii/S0927025619300138}

\bibitem{Arroyave2006473}
R.~Arroyave, A.~{van de Walle}, Z.-K. Liu,
  \href{http://www.sciencedirect.com/science/article/pii/S1359645405005574}{First-principles
  calculations of the zn–zr system}, Acta Materialia 54~(2) (2006) 473 --
  482.
\newblock \href
  {http://dx.doi.org/https://doi.org/10.1016/j.actamat.2005.09.018}
  {\path{doi:https://doi.org/10.1016/j.actamat.2005.09.018}}.
\newline\urlprefix\url{http://www.sciencedirect.com/science/article/pii/S1359645405005574}

\bibitem{Adjaoud}
O.~Adjaoud, G.~Steinle-Neumann, B.~P. Burton, A.~van~de Walle,
  \href{https://link.aps.org/doi/10.1103/PhysRevB.80.134112}{First-principles
  phase diagram calculations for the hfc--tic, zrc--tic, and hfc--zrc solid
  solutions}, Phys. Rev. B 80 (2009) 134112.
\newblock \href {http://dx.doi.org/10.1103/PhysRevB.80.134112}
  {\path{doi:10.1103/PhysRevB.80.134112}}.
\newline\urlprefix\url{https://link.aps.org/doi/10.1103/PhysRevB.80.134112}

\bibitem{ghosh:sqsce}
G.~Ghosh, A.~van~de Walle, M.~D. Asta, First-principles calculations of
  properties of bcc, fcc and hcp solid solutio\ ns in {Al-TM} ({TM = Ti, Zr,
  Hf}) systems: A comparison between cluster expansion and supercell methods,
  Acta Mater. 56 (2008) 3202.
\newblock \href {http://dx.doi.org/10.1016/j.actamat.2008.03.006}
  {\path{doi:10.1016/j.actamat.2008.03.006}}.

\bibitem{ghosh:hfnb}
G.~Ghosh, A.~van~de Walle, M.~D. Asta, G.~Olson, Phase stability of the {Hf-Nb}
  system: From first-principles to {CALPHAD}, Calphad 26 (2002) 491.
\newblock \href {http://dx.doi.org/10.1016/S0364-5916(02)80003-7}
  {\path{doi:10.1016/S0364-5916(02)80003-7}}.

\bibitem{ghosh:AlZnTi}
G.~Ghosh, A.~van~de Walle, M.~D. Asta, First-principles phase stability
  calculations of pseudobinary alloys of {(Al,Zn)$_3$Ti} with {L1$_2$},
  {DO$_{22}$} and {DO$_{23}$} structu\ res, J. Phase Equilib. Diff. 28 (2007)
  9.
\newblock \href {http://dx.doi.org/10.1007/s11669-006-9007-4}
  {\path{doi:10.1007/s11669-006-9007-4}}.

\bibitem{mediukh2019101643}
N.~Mediukh, V.~Ivashchenko, P.~Turchi, V.~Shevchenko, J.~Leszczynski, L.~Gorb,
  \href{http://www.sciencedirect.com/science/article/pii/S0364591619300719}{Phase
  diagrams and mechanical properties of tic-sic solid solutions from
  first-principles}, Calphad 66 (2019) 101643.
\newblock \href
  {http://dx.doi.org/https://doi.org/10.1016/j.calphad.2019.101643}
  {\path{doi:https://doi.org/10.1016/j.calphad.2019.101643}}.
\newline\urlprefix\url{http://www.sciencedirect.com/science/article/pii/S0364591619300719}

\bibitem{wang201955}
R.~Wang, X.~Zhang, H.~Wang, J.~Ni,
  \href{http://www.sciencedirect.com/science/article/pii/S0364591618301548}{Phase
  diagrams and elastic properties of the fe-cr-al alloys: A first-principles
  based study}, Calphad 64 (2019) 55 -- 65.
\newblock \href
  {http://dx.doi.org/https://doi.org/10.1016/j.calphad.2018.11.010}
  {\path{doi:https://doi.org/10.1016/j.calphad.2018.11.010}}.
\newline\urlprefix\url{http://www.sciencedirect.com/science/article/pii/S0364591618301548}

\bibitem{burton2006222}
B.~Burton, A.~{van de Walle},
  \href{http://www.sciencedirect.com/science/article/pii/S0009254105003554}{First-principles
  phase diagram calculations for the system nacl–kcl: The role of excess
  vibrational entropy}, Chemical Geology 225~(3) (2006) 222 -- 229, solid
  solutions: from theory to experiment.
\newblock \href
  {http://dx.doi.org/https://doi.org/10.1016/j.chemgeo.2005.08.016}
  {\path{doi:https://doi.org/10.1016/j.chemgeo.2005.08.016}}.
\newline\urlprefix\url{http://www.sciencedirect.com/science/article/pii/S0009254105003554}

\bibitem{sqs2tdb}
A.~{van de Walle}, R.~Sun, Q.-J. Hong, S.~Kadkhodaei,
  \href{http://www.sciencedirect.com/science/article/pii/S0364591617300305}{Software
  tools for high-throughput calphad from first-principles data}, Calphad 58
  (2017) 70 -- 81.
\newblock \href
  {http://dx.doi.org/https://doi.org/10.1016/j.calphad.2017.05.005}
  {\path{doi:https://doi.org/10.1016/j.calphad.2017.05.005}}.
\newline\urlprefix\url{http://www.sciencedirect.com/science/article/pii/S0364591617300305}

\bibitem{zunger.65.353}
A.~Zunger, S.-H. Wei, L.~G. Ferreira, J.~E. Bernard,
  \href{https://link.aps.org/doi/10.1103/PhysRevLett.65.353}{Special
  quasirandom structures}, Phys. Rev. Lett. 65 (1990) 353--356.
\newblock \href {http://dx.doi.org/10.1103/PhysRevLett.65.353}
  {\path{doi:10.1103/PhysRevLett.65.353}}.
\newline\urlprefix\url{https://link.aps.org/doi/10.1103/PhysRevLett.65.353}

\bibitem{mcsqs201313}
A.~{van de Walle}, P.~Tiwary, M.~{de Jong}, D.~Olmsted, M.~Asta, A.~Dick,
  D.~Shin, Y.~Wang, L.-Q. Chen, Z.-K. Liu,
  \href{http://www.sciencedirect.com/science/article/pii/S0364591613000540}{Efficient
  stochastic generation of special quasirandom structures}, Calphad 42 (2013)
  13 -- 18.
\newblock \href
  {http://dx.doi.org/https://doi.org/10.1016/j.calphad.2013.06.006}
  {\path{doi:https://doi.org/10.1016/j.calphad.2013.06.006}}.
\newline\urlprefix\url{http://www.sciencedirect.com/science/article/pii/S0364591613000540}

\bibitem{thermoCalc1985153}
B.~Sundman, B.~Jansson, J.-O. Andersson,
  \href{http://www.sciencedirect.com/science/article/pii/0364591685900215}{The
  thermo-calc databank system}, Calphad 9~(2) (1985) 153 -- 190.
\newblock \href
  {http://dx.doi.org/https://doi.org/10.1016/0364-5916(85)90021-5}
  {\path{doi:https://doi.org/10.1016/0364-5916(85)90021-5}}.
\newline\urlprefix\url{http://www.sciencedirect.com/science/article/pii/0364591685900215}

\bibitem{FactSage2002189}
C.~Bale, P.~Chartrand, S.~Degterov, G.~Eriksson, K.~Hack, R.~{Ben Mahfoud},
  J.~Melançon, A.~Pelton, S.~Petersen,
  \href{http://www.sciencedirect.com/science/article/pii/S0364591602000354}{Factsage
  thermochemical software and databases}, Calphad 26~(2) (2002) 189 -- 228.
\newblock \href
  {http://dx.doi.org/https://doi.org/10.1016/S0364-5916(02)00035-4}
  {\path{doi:https://doi.org/10.1016/S0364-5916(02)00035-4}}.
\newline\urlprefix\url{http://www.sciencedirect.com/science/article/pii/S0364591602000354}

\bibitem{FactSage2009295}
C.~Bale, E.~Bélisle, P.~Chartrand, S.~Decterov, G.~Eriksson, K.~Hack, I.-H.
  Jung, Y.-B. Kang, J.~Melançon, A.~Pelton, C.~Robelin, S.~Petersen,
  \href{http://www.sciencedirect.com/science/article/pii/S0364591608000965}{Factsage
  thermochemical software and databases — recent developments}, Calphad
  33~(2) (2009) 295 -- 311, tools for Computational Thermodynamics.
\newblock \href
  {http://dx.doi.org/https://doi.org/10.1016/j.calphad.2008.09.009}
  {\path{doi:https://doi.org/10.1016/j.calphad.2008.09.009}}.
\newline\urlprefix\url{http://www.sciencedirect.com/science/article/pii/S0364591608000965}

\bibitem{pandat2009328}
W.~Cao, S.-L. Chen, F.~Zhang, K.~Wu, Y.~Yang, Y.~Chang, R.~Schmid-Fetzer,
  W.~Oates,
  \href{http://www.sciencedirect.com/science/article/pii/S0364591608000709}{Pandat
  software with panengine, panoptimizer and panprecipitation for
  multi-component phase diagram calculation and materials property simulation},
  Calphad 33~(2) (2009) 328 -- 342, tools for Computational Thermodynamics.
\newblock \href
  {http://dx.doi.org/https://doi.org/10.1016/j.calphad.2008.08.004}
  {\path{doi:https://doi.org/10.1016/j.calphad.2008.08.004}}.
\newline\urlprefix\url{http://www.sciencedirect.com/science/article/pii/S0364591608000709}

\bibitem{OpenCalphad2015}
B.~Sundman, U.~R. Kattner, M.~Palumbo, S.~G. Fries,
  \href{https://doi.org/10.1186/s40192-014-0029-1}{Opencalphad - a free
  thermodynamic software}, Integrating Materials and Manufacturing Innovation
  4~(1) (2015) 1--15.
\newblock \href {http://dx.doi.org/10.1186/s40192-014-0029-1}
  {\path{doi:10.1186/s40192-014-0029-1}}.
\newline\urlprefix\url{https://doi.org/10.1186/s40192-014-0029-1}

\bibitem{wei2020119848}
C.~Wei, Y.~Liu, X.~Zhu, X.~Chen, Y.~Zhou, G.~Yuan, Y.~Gong, J.~Liu,
  \href{http://www.sciencedirect.com/science/article/pii/S0142961220300946}{Iridium/ruthenium
  nanozyme reactors with cascade catalytic ability for synergistic oxidation
  therapy and starvation therapy in the treatment of breast cancer},
  Biomaterials 238 (2020) 119848.
\newblock \href
  {http://dx.doi.org/https://doi.org/10.1016/j.biomaterials.2020.119848}
  {\path{doi:https://doi.org/10.1016/j.biomaterials.2020.119848}}.
\newline\urlprefix\url{http://www.sciencedirect.com/science/article/pii/S0142961220300946}

\bibitem{vukovic1992663}
M.~Vuković, D.~Čukman, M.~Milun, L.~D. Atanasoska, R.~T. Atanasoski,
  \href{http://www.sciencedirect.com/science/article/pii/0022072892803352}{Anodic
  stability and electrochromism of electrodeposited ruthenium-iridium coatings
  on titanium}, Journal of Electroanalytical Chemistry 330~(1) (1992) 663 --
  673, an International Journal Devoted to all Aspects of Electrode Kinetics,
  Interfacial Structure, Properties of Electrolytes, Colloid and Biological
  Electrochemistry.
\newblock \href
  {http://dx.doi.org/https://doi.org/10.1016/0022-0728(92)80335-2}
  {\path{doi:https://doi.org/10.1016/0022-0728(92)80335-2}}.
\newline\urlprefix\url{http://www.sciencedirect.com/science/article/pii/0022072892803352}

\bibitem{shan2019445}
J.~Shan, C.~Guo, Y.~Zhu, S.~Chen, L.~Song, M.~Jaroniec, Y.~Zheng, S.-Z. Qiao,
  \href{http://www.sciencedirect.com/science/article/pii/S2451929418305278}{Charge-redistribution-enhanced
  nanocrystalline ru@irox electrocatalysts for oxygen evolution in acidic
  media}, Chem 5~(2) (2019) 445 -- 459.
\newblock \href
  {http://dx.doi.org/https://doi.org/10.1016/j.chempr.2018.11.010}
  {\path{doi:https://doi.org/10.1016/j.chempr.2018.11.010}}.
\newline\urlprefix\url{http://www.sciencedirect.com/science/article/pii/S2451929418305278}

\bibitem{Ktz_1985}
R.~Kötz, S.~Stucki, \href{https://doi.org/10.1149%2F1.2113735}{Oxygen
  evolution and corrosion on ruthenium-iridium alloys}, Journal of The
  Electrochemical Society 132~(1) (1985) 103--107.
\newblock \href {http://dx.doi.org/10.1149/1.2113735}
  {\path{doi:10.1149/1.2113735}}.
\newline\urlprefix\url{https://doi.org/10.1149%2F1.2113735}

\bibitem{hart:PGM}
G.~L.~W. Hart, S.~Curtarolo, T.~B. Massalski, O.~Levy,
  \href{https://link.aps.org/doi/10.1103/PhysRevX.3.041035}{Comprehensive
  search for new phases and compounds in binary alloy systems based on
  platinum-group metals, using a computational first-principles approach},
  Phys. Rev. X 3 (2013) 041035.
\newblock \href {http://dx.doi.org/10.1103/PhysRevX.3.041035}
  {\path{doi:10.1103/PhysRevX.3.041035}}.
\newline\urlprefix\url{https://link.aps.org/doi/10.1103/PhysRevX.3.041035}

\bibitem{avdw:resubpd}
A.~van~de Walle, J.~Sabisch, A.~M. Minor, M.~D. Asta, {{}}{I}dentifying rhenium
  substitute candidate multi-principal-element alloys from electronic structure
  and thermodynamic criteria, Journal of Materials Research 34 (2019) 3296.
\newblock \href {http://dx.doi.org/10.1557/jmr.2019.179}
  {\path{doi:10.1557/jmr.2019.179}}.

\bibitem{l122004635}
Y.~Himuro, Y.~Tanaka, N.~Kamiya, I.~Ohnuma, R.~Kainuma, K.~Ishida,
  \href{http://www.sciencedirect.com/science/article/pii/S0966979504000949}{Stability
  of ordered l12 phase in ni3fe–ni3x (x:si and al) pseudobinary alloys},
  Intermetallics 12~(6) (2004) 635 -- 643.
\newblock \href
  {http://dx.doi.org/https://doi.org/10.1016/j.intermet.2004.03.008}
  {\path{doi:https://doi.org/10.1016/j.intermet.2004.03.008}}.
\newline\urlprefix\url{http://www.sciencedirect.com/science/article/pii/S0966979504000949}

\bibitem{D019Kang2001}
S.-Y. Kang, H.~Onodera,
  \href{https://doi.org/10.1361/105497101770332983}{Analyses of hcp/d019 and
  d019/l10 phase boundaries in ti-al-x (x=v, mn, nb, cr, mo, ni, and co)
  systems by the cluster variation method}, Journal of Phase Equilibria 22~(4)
  (2001) 424--430.
\newblock \href {http://dx.doi.org/10.1361/105497101770332983}
  {\path{doi:10.1361/105497101770332983}}.
\newline\urlprefix\url{https://doi.org/10.1361/105497101770332983}

\bibitem{PAW_PhysRevB.50.17953}
P.~E. Bl\"ochl,
  \href{https://link.aps.org/doi/10.1103/PhysRevB.50.17953}{Projector
  augmented-wave method}, Phys. Rev. B 50 (1994) 17953--17979.
\newblock \href {http://dx.doi.org/10.1103/PhysRevB.50.17953}
  {\path{doi:10.1103/PhysRevB.50.17953}}.
\newline\urlprefix\url{https://link.aps.org/doi/10.1103/PhysRevB.50.17953}

\bibitem{PBE_PhysRevLett.77.3865}
J.~P. Perdew, K.~Burke, M.~Ernzerhof,
  \href{https://link.aps.org/doi/10.1103/PhysRevLett.77.3865}{Generalized
  gradient approximation made simple}, Phys. Rev. Lett. 77 (1996) 3865--3868.
\newblock \href {http://dx.doi.org/10.1103/PhysRevLett.77.3865}
  {\path{doi:10.1103/PhysRevLett.77.3865}}.
\newline\urlprefix\url{https://link.aps.org/doi/10.1103/PhysRevLett.77.3865}

\bibitem{vasp_kresse199615}
G.~Kresse, J.~Furthmüller,
  \href{http://www.sciencedirect.com/science/article/pii/0927025696000080}{Efficiency
  of ab-initio total energy calculations for metals and semiconductors using a
  plane-wave basis set}, Computational Materials Science 6~(1) (1996) 15 -- 50.
\newblock \href
  {http://dx.doi.org/https://doi.org/10.1016/0927-0256(96)00008-0}
  {\path{doi:https://doi.org/10.1016/0927-0256(96)00008-0}}.
\newline\urlprefix\url{http://www.sciencedirect.com/science/article/pii/0927025696000080}

\bibitem{vasp_PhysRevB.54.11169}
G.~Kresse, J.~Furthm\"uller,
  \href{https://link.aps.org/doi/10.1103/PhysRevB.54.11169}{Efficient iterative
  schemes for ab initio total-energy calculations using a plane-wave basis
  set}, Phys. Rev. B 54 (1996) 11169--11186.
\newblock \href {http://dx.doi.org/10.1103/PhysRevB.54.11169}
  {\path{doi:10.1103/PhysRevB.54.11169}}.
\newline\urlprefix\url{https://link.aps.org/doi/10.1103/PhysRevB.54.11169}

\bibitem{CG10.5555/148286}
W.~H. Press, S.~A. Teukolsky, W.~T. Vetterling, B.~P. Flannery, Numerical
  Recipes in C (2nd Ed.): The Art of Scientific Computing, Cambridge University
  Press, USA, 1992.

\bibitem{MPPhysRevB.40.3616}
M.~Methfessel, A.~T. Paxton,
  \href{https://link.aps.org/doi/10.1103/PhysRevB.40.3616}{High-precision
  sampling for brillouin-zone integration in metals}, Phys. Rev. B 40 (1989)
  3616--3621.
\newblock \href {http://dx.doi.org/10.1103/PhysRevB.40.3616}
  {\path{doi:10.1103/PhysRevB.40.3616}}.
\newline\urlprefix\url{https://link.aps.org/doi/10.1103/PhysRevB.40.3616}

\bibitem{BlochlPhysRevB.49.16223}
P.~E. Bl\"ochl, O.~Jepsen, O.~K. Andersen,
  \href{https://link.aps.org/doi/10.1103/PhysRevB.49.16223}{Improved
  tetrahedron method for brillouin-zone integrations}, Phys. Rev. B 49 (1994)
  16223--16233.
\newblock \href {http://dx.doi.org/10.1103/PhysRevB.49.16223}
  {\path{doi:10.1103/PhysRevB.49.16223}}.
\newline\urlprefix\url{https://link.aps.org/doi/10.1103/PhysRevB.49.16223}

\bibitem{AIMDK_hne_2014}
T.~D. Kühne, \href{http://dx.doi.org/10.1002/wcms.1176}{Second generation
  car-parrinello molecular dynamics}, Wiley Interdisciplinary Reviews:
  Computational Molecular Science 4~(4) (2014) 391–406.
\newblock \href {http://dx.doi.org/10.1002/wcms.1176}
  {\path{doi:10.1002/wcms.1176}}.
\newline\urlprefix\url{http://dx.doi.org/10.1002/wcms.1176}

\bibitem{Langevindoi:10.1063/1.445195}
D.~J. Evans, \href{https://doi.org/10.1063/1.445195}{Computer
  ‘‘experiment’’ for nonlinear thermodynamics of couette flow}, The
  Journal of Chemical Physics 78~(6) (1983) 3297--3302.
\newblock \href {http://arxiv.org/abs/https://doi.org/10.1063/1.445195}
  {\path{arXiv:https://doi.org/10.1063/1.445195}}, \href
  {http://dx.doi.org/10.1063/1.445195} {\path{doi:10.1063/1.445195}}.
\newline\urlprefix\url{https://doi.org/10.1063/1.445195}

\bibitem{LangevinPhysRevLett.48.1818}
W.~G. Hoover, A.~J.~C. Ladd, B.~Moran,
  \href{https://link.aps.org/doi/10.1103/PhysRevLett.48.1818}{High-strain-rate
  plastic flow studied via nonequilibrium molecular dynamics}, Phys. Rev. Lett.
  48 (1982) 1818--1820.
\newblock \href {http://dx.doi.org/10.1103/PhysRevLett.48.1818}
  {\path{doi:10.1103/PhysRevLett.48.1818}}.
\newline\urlprefix\url{https://link.aps.org/doi/10.1103/PhysRevLett.48.1818}

\bibitem{fitfc_2009}
A.~van~de Walle,
  \href{http://www.sciencedirect.com/science/article/pii/S0364591608001314}{Multicomponent
  multisublattice alloys, nonconfigurational entropy and other additions to the
  {Alloy} {Theoretic} {Automated} {Toolkit}}, Calphad 33~(2) (2009) 266--278.
\newblock \href {http://dx.doi.org/10.1016/j.calphad.2008.12.005}
  {\path{doi:10.1016/j.calphad.2008.12.005}}.
\newline\urlprefix\url{http://www.sciencedirect.com/science/article/pii/S0364591608001314}

\bibitem{vandewall_SGTE20181}
A.~{van de Walle},
  \href{http://www.sciencedirect.com/science/article/pii/S0364591617301505}{Invited
  paper: Reconciling sgte and ab initio enthalpies of the elements}, Calphad 60
  (2018) 1 -- 6.
\newblock \href
  {http://dx.doi.org/https://doi.org/10.1016/j.calphad.2017.10.008}
  {\path{doi:https://doi.org/10.1016/j.calphad.2017.10.008}}.
\newline\urlprefix\url{http://www.sciencedirect.com/science/article/pii/S0364591617301505}

\bibitem{mech_vandeWalle2015}
A.~van~de Walle, Q.~Hong, S.~Kadkhodaei, R.~Sun,
  \href{https://doi.org/10.1038/ncomms8559}{The free energy of mechanically
  unstable phases}, Nature Communications 6~(1) (2015) 7559.
\newblock \href {http://dx.doi.org/10.1038/ncomms8559}
  {\path{doi:10.1038/ncomms8559}}.
\newline\urlprefix\url{https://doi.org/10.1038/ncomms8559}

\bibitem{ansara199720}
I.~Ansara, N.~Dupin, H.~L. Lukas, B.~Sundman,
  \href{http://www.sciencedirect.com/science/article/pii/S0925838896026527}{Thermodynamic
  assessment of the al-ni system}, Journal of Alloys and Compounds 247~(1)
  (1997) 20 -- 30.
\newblock \href
  {http://dx.doi.org/https://doi.org/10.1016/S0925-8388(96)02652-7}
  {\path{doi:https://doi.org/10.1016/S0925-8388(96)02652-7}}.
\newline\urlprefix\url{http://www.sciencedirect.com/science/article/pii/S0925838896026527}

\bibitem{IrRuphase1}
K.~V. Yusenko, S.~A. Martynova, S.~Khandarkhaeva, T.~Fedotenko, K.~Glazyrin,
  E.~Koemets, M.~Bykov, M.~Hanfland, K.~Siemensmeyer, A.~Smekhova, S.~A.
  Gromilov, L.~S. Dubrovinsky,
  \href{http://www.sciencedirect.com/science/article/pii/S2589152920303367}{High
  compressibility of synthetic analogous of binary iridium–ruthenium and
  ternary iridium–osmium–ruthenium minerals}, Materialia 14 (2020) 100920.
\newblock \href {http://dx.doi.org/https://doi.org/10.1016/j.mtla.2020.100920}
  {\path{doi:https://doi.org/10.1016/j.mtla.2020.100920}}.
\newline\urlprefix\url{http://www.sciencedirect.com/science/article/pii/S2589152920303367}

\bibitem{IrRuphase2}
N.~A. Saltykova, O.~V. Portnyagin,
  \href{https://doi.org/10.1023/A:1011944226271}{Electrodeposition of
  iridium--ruthenium alloys from chloride melts: the structure of the
  deposits}, Russian Journal of Electrochemistry 37~(9) (2001) 924--930.
\newblock \href {http://dx.doi.org/10.1023/A:1011944226271}
  {\path{doi:10.1023/A:1011944226271}}.
\newline\urlprefix\url{https://doi.org/10.1023/A:1011944226271}

\bibitem{okamoto_pd}
H.~Okamoto, Binary alloy phase diagrams, Vol.~3, ASM International, 1990.

\bibitem{sro:kikuchi}
R.~Kikuchi, \href{https://link.aps.org/doi/10.1103/PhysRev.81.988}{A theory of
  cooperative phenomena}, Phys. Rev. 81 (1951) 988--1003.
\newblock \href {http://dx.doi.org/10.1103/PhysRev.81.988}
  {\path{doi:10.1103/PhysRev.81.988}}.
\newline\urlprefix\url{https://link.aps.org/doi/10.1103/PhysRev.81.988}

\bibitem{sro:Pelizzola_2005}
A.~Pelizzola, \href{http://dx.doi.org/10.1088/0305-4470/38/33/R01}{Cluster
  variation method in statistical physics and probabilistic graphical models},
  Journal of Physics A: Mathematical and General 38~(33) (2005) R309–R339.
\newblock \href {http://dx.doi.org/10.1088/0305-4470/38/33/r01}
  {\path{doi:10.1088/0305-4470/38/33/r01}}.
\newline\urlprefix\url{http://dx.doi.org/10.1088/0305-4470/38/33/R01}

\bibitem{sro:kkr-cpa1}
S.~N. Khan, J.~B. Staunton, G.~M. Stocks,
  \href{https://link.aps.org/doi/10.1103/PhysRevB.93.054206}{Statistical
  physics of multicomponent alloys using kkr-cpa}, Phys. Rev. B 93 (2016)
  054206.
\newblock \href {http://dx.doi.org/10.1103/PhysRevB.93.054206}
  {\path{doi:10.1103/PhysRevB.93.054206}}.
\newline\urlprefix\url{https://link.aps.org/doi/10.1103/PhysRevB.93.054206}

\bibitem{sro:johnson}
D.~D. Johnson, D.~M. Nicholson, F.~J. Pinski, B.~L. Gy\"orffy, G.~M. Stocks,
  \href{https://link.aps.org/doi/10.1103/PhysRevB.41.9701}{Total-energy and
  pressure calculations for random substitutional alloys}, Phys. Rev. B 41
  (1990) 9701--9716.
\newblock \href {http://dx.doi.org/10.1103/PhysRevB.41.9701}
  {\path{doi:10.1103/PhysRevB.41.9701}}.
\newline\urlprefix\url{https://link.aps.org/doi/10.1103/PhysRevB.41.9701}

\bibitem{sro:cpa_soven}
P.~Soven,
  \href{https://link.aps.org/doi/10.1103/PhysRev.156.809}{Coherent-potential
  model of substitutional disordered alloys}, Phys. Rev. 156 (1967) 809--813.
\newblock \href {http://dx.doi.org/10.1103/PhysRev.156.809}
  {\path{doi:10.1103/PhysRev.156.809}}.
\newline\urlprefix\url{https://link.aps.org/doi/10.1103/PhysRev.156.809}

\bibitem{sro:Zeller_1987}
R.~Zeller,
  \href{https://doi.org/10.1088/0022-3719/20/16/010}{Multiple-scattering
  solution of schrodinger{\textquotesingle}s equation for potentials of general
  shape}, Journal of Physics C: Solid State Physics 20~(16) (1987) 2347--2360.
\newblock \href {http://dx.doi.org/10.1088/0022-3719/20/16/010}
  {\path{doi:10.1088/0022-3719/20/16/010}}.
\newline\urlprefix\url{https://doi.org/10.1088/0022-3719/20/16/010}

\end{thebibliography}







\end{document}